%
\magnification = \magstep1
\hsize = 16 truecm 
\parskip = 0pt
\normalbaselineskip = 18pt plus 0.2pt minus 0.1pt
\baselineskip = \normalbaselineskip
%
%

\font\brm=cmr10 scaled \magstep1
\font\bbf=cmbx10 scaled \magstep1
\font\bit=cmti10 scaled \magstep1



%

\def\Bigbreak{\par \ifdim\lastskip < \bigskipamount \removelastskip \fi
                   \penalty-300 \vskip 10mm plus 5mm minus 2mm}
%
%
\newcount\eqnum
\newcount\tempeq
\def\cleareqnum{\global\eqnum = 0}
\def\eqname{(\the\chnum.\the\eqnum)}
\def\neweq{\global\advance\eqnum by 1 \eqno\eqname}
\def\neweqalign{\global\advance\eqnum by 1 &\eqname}
\def\releq#1{\global\tempeq=\eqnum \advance\tempeq by #1
             (\the\chnum.\the\tempeq)}

\cleareqnum
%
%
\newcount\chnum
\def\clearchnum{\global\chnum = 0}
%
%

\def\newchapt#1{\Bigbreak \global\advance\chnum by 1
                \cleareqnum
                \centerline{\bbf\the\chnum.{ }#1}
                \nobreak\vskip 5mm plus 2mm minus 1mm}
\clearchnum
%
%
\newcount\notenumber
\def\clearnotenumber{\notenumber=0} 
\def\note{\advance\notenumber by1 \footnote{$^{\the\notenumber}$}} 
\clearnotenumber 
%
%
\newbox\Eqa
\newbox\Eqb
\newbox\Eqc
\newbox\Eqd
\newbox\Eqe
\newbox\Eqf
\newbox\Eqg
\newbox\Eqh
\newbox\Eqi
\def\storeeq#1{\setbox #1=\hbox{\eqname}}

%
%
\pageno=0
\footline={\ifnum\pageno=0\strut\hfil\else\hfil\tenrm\folio\hfil\strut\fi}%
\def\e{{\rm e}}
\def\H{{\cal H}}

\def\E{{\cal E}}
\def\F{{\cal F}}

\def\V{{\cal V}}

\def\S{{\cal S}}

\def\A{{\cal A}}

\def\Q{{\cal Q}}
\def\cZ{{\cal Z}}
\def\R{{\bf R}}
\def\C{{\bf C}} 
\def\Z{{\bf Z}} 
\def\N{{\bf N}} 
\def\der{\partial } 
\def\mis{{dk\over 2\pi }} 
\def\sc{\varphi } 
\def\dsc{\widetilde \varphi} 
\def\scr{\varphi_{{}_R}} 
\def\scl{\varphi_{{}_L}}
\def\spr{\psi_{{}_R}} 
\def\spl{\psi_{{}_L}}
\def\psc{\varphi_\lambda } 
\def\pdsc{{\widetilde \varphi}_\lambda } 

\def\alg{{\cal A}} 
\def\palg{{\cal A}_\lambda }
\def\form{\langle \, \cdot \, , \, \cdot \, \rangle } 
\def\sform{(\, \cdot \, ,\, \cdot \, )}
\def\hl{\overline \R_+} 
%
%

\rightline {IFUP-TH 26/97} 

\rightline {ref. SISSA 84/97/FM}

\vskip 1.5 truecm
\centerline {\bbf Quantum Field Theory, Bosonization and} 
\centerline {\bbf Duality on the Half Line}
\vskip 1 truecm
\centerline {\brm Antonio Liguori} 
\medskip
\centerline {\it International School for Advanced Studies,}
\centerline {\it 34014 Trieste, Italy}
\bigskip
\medskip  
\centerline {\brm Mihail Mintchev} 
\medskip
\centerline {\it Istituto Nazionale di Fisica Nucleare, Sezione di Pisa}
\centerline {\it Dipartimento di Fisica dell'Universit\`a di Pisa,}
\centerline {\it Piazza Torricelli 2, 56100 Pisa, Italy}
\bigskip 
\vskip 2 truecm
\centerline {\bit Abstract} 
\medskip

We develop a bosonization procedure on the half line. Different boundary 
conditions, formulated in terms of the vector and axial fermion currents, 
are implemented by using in general the mixed boundary condition on the 
bosonic field. The interplay between symmetries and boundary conditions is 
investigated in this context, with a particular emphasis on duality. 
As an application, we explicitly construct operator solutions of the massless 
Thirring model on the half line, respecting different boundary conditions.


\bigskip 
\bigskip  
\centerline {September 1997} 
\vfill \eject 

\newchapt {Introduction} 

Quantum field theories in space-time with boundaries are useful 
in different areas of physics, providing deeper understanding 
of various boundary phenomena. The renewed interest 
in the subject stems from the recent advance [1-4] in 
the treatment of impurity problems from condensed matter 
physics and is related also to applications in dissipative quantum 
mechanics [5] and open string theory [6]. In the simplest cases, one 
is dealing with models in 1+1 dimensions, constructed in terms of fields 
which are free everywhere except for an interaction on the boundary. 
In spite of the remarkable progress achieved in handling these boundary 
interactions by means of conformal and/or integrable field 
theory [7-12], some interesting issues need a further investigation. One may wonder 
in this respect if the method of bosonization, 
another fundamental tool of two-dimensional quantum field theory, can shed new 
light on the subject. In absence of boundaries, bosonization plays an 
essential role for constructing exact solutions [13] of some 
models and establishing the equivalence among others [14,15]. 
The nonabelian variant of bosonization [16] produces also relevant results 
in both conformal field and string theory. It is quite natural at this point 
to try bosonization in 1+1 dimensional space-time with boundaries. 
To our knowledge however, a systematic treatment is still to be developed 
in this case, where both translation and Lorentz invariance are broken and the 
deeply related concepts of locality and statistics have to 
be reconsidered. Together with the notion of duality, these are 
the basic problems which must be faced in order to 
develop bosonization with boundaries. 

In the present paper we propose a bosonization procedure 
on the half line\break $\hl = \{x\in \R\, :\, x\geq 0\}$, 
addressing the just mentioned problems. Our work is organized as follows. 
We start by investigating the free  massless 
scalar $\sc $ field with mixed boundary condition (in $x=0$), fixed 
in terms of the so called boundary parameter $\eta $. Afterwards, we 
construct the dual field $\dsc $ and establish the locality properties of 
$\sc $ and $\dsc $, surviving the breakdown of the spatial 
symmetries on $\R \times \hl $. These locality properties 
are the basis of bosonization. We also 
compute the correlation functions of $\sc $ and $\dsc $, detecting 
the absence of left-right factorization. As expected, the symmetry 
content of the fields $\sc $ and $\dsc $ depends on the value of the 
parameter $\eta $. It turns out in particular, that duality connects the 
Neumann and Dirichlet boundary conditions. All these aspects are discussed in 
detail in section 2. In section 3 we build anyon fields 
in terms of $\sc $ and $\dsc $, elaborating on the delicate 
question concerning the metric in the state space. Section 4 is 
devoted to the free fermion bosonization. We describe various 
types of boundary conditions, which are most naturally formulated 
in terms of the vector and/or the axial current and are therefore 
nonlinear in the fermion field. The deep relationship between internal 
symmetries and boundary conditions is also investigated. 
In section 5 we turn to interacting fields, considering the 
Thirring model on $\R \times \hl $ in three 
different phases, corresponding to what we call vector, axial 
and vector-axial symmetric boundary conditions. In order to construct the conserved 
currents, we examine the short distance behavior of the corresponding operator 
products. We discuss the subtle points related to the absence of 
Lorentz and translation invariance and explicitly derive the renormalization 
constants needed in defining the vector and axial currents.  The last section 
is devoted to our conclusions. Some technical issues are collected in the appendices.

\bigskip

\newchapt {The Massless Scalar Field and Its Dual on The Half Line}

The idea for bosonization can be traced back to 
Jordan and Wigner [17]. From the rather extensive literature on the subject 
(see e.g. [18,19] and references therein), it is well known 
that the fundamental building blocks in $\R \times \R$ 
are the free massless scalar field 
$\sc $ and its dual $\dsc $. These fields are related via  
$$
\der_x \dsc (t,x) = - \der_t \sc (t,x) \quad , \qquad 
\der_t \dsc (t,x) = - \der_x \sc (t,x) \quad . \neweq
$$
Both $\sc $ and $\dsc $ are {\it local} fields, but it turns out 
that they are not {\it relatively} local. Therefore, the 
Wick monomials in $\sc $ and $\dsc $ respectively, generate 
two different Borchers classes [20]. 
As recognized already in the early sixties [18], the deep 
interplay between these two classes is in the heart 
of bosonization, duality, chiral superselection sectors 
and many other phenomena, characterizing the rich 
and fascinating structure of 1+1 dimensional quantum field 
theory. For this reason, it is natural to ask to what 
extent the above features of the fields $\sc $ and $\dsc $ 
hold also in the presence of a boundary. For answering 
this question, we investigate in this section  
the problem on $\R \times \hl $. 

So, let us start with the classical action 
$$
S[\sc ] = {1\over 2}\int_{-\infty}^\infty dt \int_0^\infty dx 
\left \{ \left [\der_t \sc (t,x)\right ]^2 - 
\left [\der_x \sc (t,x)\right ]^2 \right \}
-{\eta\over 2}\int_{-\infty}^\infty dt \, \varphi^2(t,0)
\quad, \neweq
$$
which has the usual ``bulk'' part plus a boundary term. 
Varying $S[\sc ]$ one gets the standard equation of motion 
$$ 
(\der_t^2 - \der_x^2) \sc (t,x) = 0 \quad , \neweq 
$$
and the boundary condition 
$$
\lim_{x \downarrow 0}\left( \der_x - \eta \right) 
\sc (t,x) = 0 \quad . \neweq 
$$
Eq.(2.4) represents the so called mixed boundary 
condition. Notice that $\eta $ has dimension of mass. 
In the limits $\eta \to 0$ and $\eta \to \infty $ 
one formally recovers from (2.4) the familiar Neumann and 
Dirichlet boundary conditions respectively. 

One should always keep in mind that the boundary term 
in (2.2) breaks down the invariance under 
Lorentz boosts and space translations. $S[\sc ]$ is 
invariant however under time translations, which implies 
conservation of the energy 
$$
E[\sc ] = {1\over 2}\int_0^\infty dx \left \{ 
\left [\der_t \sc (t,x)\right ]^2 + 
\left [\der_x \sc (t,x)\right ]^2 \right \} + 
{\eta\over 2} \sc^2(t,0) \quad . 
\neweq
$$
Indeed, because of eq.(2.4), the time derivatives of the 
integral and the boundary term in (2.5) precisely compensate 
each other. From eq.(2.5) we deduce that the functional $E[\sc ]$ 
is non-negative if and only if $\eta \geq 0$, which will be 
assumed in the rest of this paper. The case $\eta < 0$ is 
also physically interesting and will be considered in 
a future publication [21]. 

Integrating by parts and using (2.4), $E[\sc ]$ can be written 
in the form 
$$
E[\sc ] = {1\over 2}\int_0^\infty dx \left \{ 
\left [\der_t \sc (t,x)\right ]^2 - 
\sc (t,x)\der_x^2 \sc (t,x) \right \}  \quad , \neweq 
$$
which implies the following energy density: 
$$
\E (t,x) = {1\over 2} \left \{ [\der_t \sc (t,x) ]^2 - 
\sc (t,x)\der_x^2 \sc (t,x)\right \} \quad . \neweq 
$$

We turn now to the construction of the quantum fields 
$\sc $ and $\dsc $. First of all, we need a complete set of 
positive energy solutions of eq.(2.3), which satisfy the 
boundary condition (2.4). For this purpose we consider 
the functions 
$$
\psi_k(t,x) = \e^{ikt}\left [
\e^{-ikx} + B(k)\, \e^{ikx} \right ]\quad , 
\quad \quad k\in \R \quad , \neweq 
$$
where 
$$
B(k) \equiv {k-i\eta \over k+i\eta} \quad . \neweq 
$$
The phase factor $B(k)$ obeys the following useful relations: 
$$
B(k)\, B(-k) = 1 \quad , \qquad {\overline B}(k)\, B(k) = 1 \quad . 
\neweq 
$$
One can directly verify that the family 
$\{\psi_k \, :\, k > 0\}$ fulfills all the requirements. 
In particular, for fixed time one has 
both orthogonality 
$$
\int_0^\infty dx\,{\overline \psi}_k(t,x) {\psi}_p(t,x) = 
2\pi\delta (k-p) 
\quad , \neweq 
$$
and completeness 
$$
\int_0^\infty \mis \,{\overline \psi}_k(t,x) {\psi}_k(t,y) = 
\delta (x-y) 
\quad , \neweq
$$ 
where the variables in (2.11,12) belong to $\R_+$. 
In order to define the dual field $\dsc $, we also introduce the set 
$$
{\widetilde \psi}_k(t,x) = \e^{ikt}\left [
\e^{-ikx} - B(k)\, \e^{ikx} \right ]\quad , 
\quad \quad k \in \R \quad , \neweq 
$$
of solutions of (2.3), which satisfy in addition the duality equations 
$$
\der_x {\widetilde \psi}_k (t,x) = - \der_t \psi_k (t,x) \quad , \qquad 
\der_t {\widetilde \psi}_k (t,x) = - \der_x \psi_k (t,x) \quad , 
\neweq
$$
analogous to (2.1). 

The last ingredient, needed for the construction of 
the fields $\sc $ and $\dsc $ on the half line, is the algebra 
$\alg $ generated by $\{a^\ast (k), a(k)\, :\, k\geq 0 \}$, 
satisfying the commutation relations 
$$
[a(k)\, ,\, a(p)] = [a^\ast (k)\, ,\, a^\ast (p)] = 0 
\quad , \neweq 
$$
$$ 
[a(k)\, ,\, a^\ast (p)] = 2\pi \left (k_+^{-1}\right )_{\mu_{{}_0}} \, \delta (k-p)  
\quad . \neweq 
$$
Here $\mu_0 > 0$ is a scale parameter with dimension of mass. 
The infrared origin of $\mu_0$, together with 
the definition and the basic properties of the distribution 
$\left (k_+^{-1}\right )_{\mu_{{}_0}}$, are discussed in Appendix A. 
There one can find also the explicit construction of the Fock 
representation $\F$ of $\alg $, used throughout the paper. Since 
$\left (k_+^{-1}\right )_{\mu_{{}_0}}$ is not positive definite, 
$\F$ has indefinite metric - a typical feature of the 
two-dimensional world [18]. 

At this stage we are ready to introduce the quantum fields 
$\sc $ and $\dsc $ defined by 
$$
\sc (t,x) = {1\over \sqrt 2} \int_0^\infty \mis 
\left [a^\ast (k) \psi_k (t,x) + 
a(k){\overline \psi}_k(t,x)\right ] \quad , \neweq 
$$
$$
\dsc (t,x) = {1\over \sqrt 2} \int_0^\infty \mis 
\left [a^\ast (k) {\widetilde \psi}_k (t,x) + 
a(k){\overline {\widetilde \psi}}_k(t,x)\right ] \quad . 
\neweq 
$$
Both $\sc $ and $\dsc $ are free massless Klein-Gordon fields, expressed 
in terms of the same creation and annihilation operators 
$\{a^\ast (k), a(k)\, :\, k\geq 0 \}$, acting in $\F$. 
The properties of $\psi_k$ and ${\widetilde \psi}_k$ imply 
that $\sc $ and $\dsc $ satisfy the the duality relation (2.1) and the 
boundary condition (2.4) with arbitrary, but fixed $\eta \geq 0$. 
Moreover, at the special values $\eta = 0$ and $\eta = \infty $ 
the phase factor (2.9) simplifies, leading to the relations 
$$
\sc (t,x)\vert_{\eta = 0} = \dsc (t,x)\vert_{\eta = \infty } 
\quad , \neweq 
$$
$$
\sc (t,x)\vert_{\eta = \infty } = \dsc (t,x)\vert_{\eta = 0 } 
\quad . \neweq 
$$
Eqs.(2.19-20) are a first manifestation 
of the nontrivial interplay between duality and boundary 
conditions on the half line.  

In order to develop the bosonization procedure we have to establish 
the commutation properties of $\sc $ and $\dsc $. The independent 
commutators can be conveniently represented in the form 
$$
[\sc (t_1,x_1)\, ,\, \sc (t_2,x_2)] = 
iD(t_1-t_2, x_1-x_2) + iD_\eta (t_1-t_2, x_1+x_2) 
\quad , \neweq 
$$
$$
[\dsc (t_1,x_1)\, ,\, \dsc (t_2,x_2)] = 
iD(t_1-t_2, x_1-x_2) - iD_\eta (t_1-t_2, x_1+x_2) 
\quad , \neweq 
$$
$$
[\sc (t_1,x_1)\, ,\, \dsc (t_2,x_2)] = 
i{\widetilde D}(t_1-t_2, x_1-x_2) + 
i{\widetilde D}_\eta (t_1-t_2, x_1+x_2) 
\quad , \neweq 
$$
where the translational non-invariant and $\eta $-dependent 
contribution has been isolated in the functions $D_\eta $ and 
${\widetilde D}_\eta $. A simple computation gives 
$$
D(t,x) = -{1\over 2} \varepsilon (t)\, \theta (t^2 - x^2) 
\quad , \neweq 
$$
$$
{\widetilde D}(t,x) = 
{1\over 2} \varepsilon (x)\, \theta (x^2 - t^2) 
\quad , \neweq 
$$
which coincide with the commutators in the absence 
of a boundary (see e.g. [18]). Using that $\eta \geq 0$, 
one also finds 
$$
D_\eta (t,x) = 
\cases { -{1\over 2} \varepsilon (t)\, \theta (t^2 - x^2) 
\, \, ,& $\eta = 0$; \cr 
{1\over 2} \varepsilon (t)\, \theta (t^2 - x^2) + 
\theta (-t-x)\e^{\eta (t+x)} - 
\theta (t-x) \e^{\eta (x-t)}\, \, , & $0<\eta <\infty $; \cr 
{1\over 2} \varepsilon (t)\, \theta (t^2 - x^2)\, \, , 
& $\eta = \infty $;\cr } \neweq 
$$ 
and 
$$
{\widetilde D}_\eta (t,x) = 
\cases { -{1\over 2} \varepsilon (x)\, \theta (x^2 - t^2) 
\, \, ,& $\eta = 0$; \cr 
{1\over 2} \varepsilon (x)\, \theta (x^2 - t^2) + 
\theta (-t-x)\e^{\eta (t+x)} + 
\theta (t-x) \e^{\eta (x-t)}\, \, , & $0<\eta <\infty $; \cr 
{1\over 2} \varepsilon (x)\, \theta (x^2 - t^2)\, \, , 
& $\eta = \infty $.\cr } \neweq 
$$
We see from eqs.(2.24-27) that the commutators (2.21-23) are 
$\mu_{{}_0}$-independent, in spite of the fact that $\sc $ and 
$\dsc $ depend on $\mu_{{}_0} $. Notice also that $D_\eta $ 
enters the right hand side of (2.21) and (2.22) with a different 
sign. Therefore, 
$$
[\sc (t_1,x_1)\, ,\, \sc (t_2,x_2)] \not= 
[\dsc (t_1,x_1)\, ,\, \dsc (t_2,x_2)] \quad , 
$$
contrary to the case without boundary [18]. Another feature 
which is worth stressing is that $D_\eta $ and 
${\widetilde D}_\eta $ are continuous in $\eta$ for 
$ 0 < \eta \le \infty $, but discontinuous in $\eta = 0$. 
We will see later on that the symmetry content of the theory also 
exhibits such a discontinuity. 

After these general remarks, we concentrate on the commutators 
(2.21,22). Observing that for $x_1,\, x_2\in {\overline \R}_+$ the 
inequality $|t_1-t_2|< |x_1-x_2|$ implies $|t_1-t_2|< x_1 + x_2$, 
one can check that $\sc $ commutes with itself at 
space-like separated points and is therefore a local field. 
This conclusion, which holds also for $\dsc $, is quite 
remarkable because Lorentz invariance is broken. 
The two terms $D$ and $D_\eta $ in (2.21,22) 
have a very intuitive explanation. As far as 
$|t_1-t_2|<|x_1-x_2|$ no signal can propagate between 
the points $(t_1,x_1)$ and $(t_2,x_2)$ and the commutator vanishes. 
When $|x_1-x_2|<|t_1-t_2|< x_1 + x_2 $ signals can propagate 
directly between the two points, but they cannot be influenced 
by the boundary and the only contribution comes from 
$D$. As soon as $ x_1 + x_2 = |t_1-t_2|$, 
signals starting from one of the points can be reflected 
at the boundary and reach the other point. This phenomenon is 
responsible for the term $D_\eta $ in eq.(2.21,22). By means of 
eqs.(2.24,26) one can also easily verify that the equal-time canonical 
commutation relations 
$$
[\sc (t,x_1)\, ,\, \sc (t,x_2)] = 
[\dsc (t,x_1)\, ,\, \dsc (t,x_2)] = 0 
\quad , \neweq 
$$
$$
[\der_t\sc (t,x_1)\, ,\, \sc (t,x_2)] = 
[\der_t\dsc (t,x_1)\, ,\, \dsc (t,x_2)] = 
-i\delta (x_1-x_2) \quad , \neweq 
$$
hold when $x_{1,2}>0$. Eq.(2.29) fixes actually the 
normalization in (2.17,18). 

Concerning relative locality, from the mixed commutator (2.23) and 
eqs.(2.25,27) we immediately see that $\sc $ is nonlocal with respect to $\dsc $. 
We will see later that this fact is of crucial importance for bosonization. 

It is both instructive and convenient to introduce also the chiral basis 
of right and left fields, defined by 
$$
\scr (t-x) \equiv [\sc (t,x) + \dsc (t,x)] = 
{\sqrt 2} \int_0^\infty \mis 
\left [a^\ast (k) \e^{ik(t-x)} + 
a(k)\e^{-ik(t-x)} \right ] 
\, \,  , \neweq 
$$
$$
\scl (t+x) \equiv [\sc (t,x) - \dsc (t,x)] = 
{\sqrt 2} \int_0^\infty \mis 
\left [a^\ast (k) B(k) \e^{ik(t+x)} + 
a(k) B(-k) \e^{-ik(t+x)}\right ]   
\, \, . \neweq 
$$
Notice that each of the fields $\scr $ and $\scl $ depends 
on one real variable - the light-cone coordinate $t-x$ and 
$t+x$ correspondingly. Using eqs.(2.21-27) one finds 
$$
[\scr (\xi_1 )\, ,\, \scr (\xi_2 )] = 
[\scl (\xi_1 )\, ,\, \scl (\xi_2 )] = 
-i\, \varepsilon (\xi_{12}) 
\quad , \qquad \xi_{12} \equiv \xi_1 - \xi_2 \quad , \neweq 
$$
which are not sensitive to the parameter $\eta $. The boundary manifests 
itself in the mixed relation 
$$
[\scr (\xi_1 )\, ,\, \scl (\xi_2 )] = \cases 
{-i\, \varepsilon (\xi_{12})\, \, , &$\eta = 0$;\cr 
i\, \varepsilon (\xi_{12}) - 
4i\theta (\xi_{12})\e^{-\eta \xi_{12}} \, \, , 
&$0<\eta < \infty $;\cr i\, \varepsilon (\xi_{12}) 
&$\eta = \infty $.\cr } \neweq 
$$
As it should be expected, due to the $\sc $-$\dsc $ mixing, 
the locality properties are less transparent in the chiral 
basis. The commutators however, have a more compact form and 
are translation invariant in the light-cone coordinates. 
We draw reader's attention to eq.(2.33). 
It shows that in contrast to the case without boundary, 
the left and the right sectors of the free massless 
scalar field on the half line do not decouple. 
The familiar left-right factorization breaks down on $\overline \R_+$ 
even for $\eta = 0$ and $\eta = \infty $. This phenomenon, 
which has relevant consequences in the applications, has been 
discussed in the context of conformal field theory by Cardy [22]. 

The fundamental correlation functions in the chiral basis are: 
$$
w_{{}_{AB}}(\xi_1, \xi_2) = 
\langle \Omega \, , \, \sc_{{}_A} (\xi_1 ) \sc_{{}_B} (\xi_2 )\Omega \rangle 
\quad , \qquad A,B = R,L \quad , \neweq 
$$
where $\Omega $ and $\form $ are the vacuum and the scalar product 
in the Fock representation $\F$ (see the appendix). Using the commutation 
relations (2.15-16) and the fact that $\Omega $ is annihilated by $a(k)$, 
we get: 
$$
w_{{}_{RR}}(\xi_1, \xi_2) = w_{{}_{LL}}(\xi_1, \xi_2) = 
u(\mu \xi_{12}) \quad , \neweq 
$$
$$
w_{{}_{RL}}(\xi_1, \xi_2) = \cases 
{u(\mu \xi_{12})\, \, , &$\eta = 0$;\cr 
-u(\mu \xi_{12}) - 
v_-(\eta \xi_{12}) \, \, , 
&$0<\eta < \infty $;\cr -u(\mu \xi_{12}) 
&$\eta = \infty $,\cr } \neweq  
$$
$$
w_{{}_{LR}}(\xi_1, \xi_2) = \cases 
{u(\mu \xi_{12})\, \, , &$\eta = 0$;\cr 
-u(\mu \xi_{12}) - 
v_+(-\eta \xi_{12}) \, \, , 
&$0<\eta < \infty $;\cr -u(\mu \xi_{12}) 
&$\eta = \infty $.\cr } \neweq  
$$
Here 
$$
\mu \equiv \mu_{{}_0} \, \e^{\gamma_{{}_E}} 
\quad , \neweq 
$$
$\gamma_{{}_E}$ being Euler's constant. 
The explicit form of the distributions $u$ and $v_\pm $ is 
$$
u(\xi ) = -{1\over \pi } \ln (i\xi + \epsilon ) 
= -{1\over \pi } \ln |\xi | - {i\over 2}\varepsilon (\xi ) 
\quad , \neweq 
$$
$$
v_\pm (\xi ) = {2\over \pi }\, \e^{-\xi }
\, {\rm Ei} (\xi \pm i\epsilon ) 
\quad , \neweq 
$$
where ${\rm Ei}$ denotes the exponential-integral function, which 
has a cut in the complex plane located on the positive real axes. 
As usual, the presence of the parameter $\epsilon >0$ implies
the weak limit $\epsilon \to 0$. One can verify that 
$$
u(\xi) - u(-\xi) = -i\varepsilon (\xi) \quad , \neweq 
$$
$$
v_-(\xi) - v_+(\xi) = 4i\theta (\xi)\e^{-\xi} \quad , 
\neweq 
$$
which provide a consistency check between (2.32,33) and 
(2.35-40).  

Now we are in the position to investigate another fundamental 
aspect - the metric in the state space 
of $\sc $ and $\dsc $. Just as in the case on the whole line, it is 
easy to show that there are states with negative norm. Indeed, let us 
consider the smeared fields 
$$
\scr (f) \equiv \int_{-\infty }^{\infty }dt \int_0^{\infty }
dx f(t,x) \scr (t-x) \quad , \qquad 
\scl (f) \equiv \int_{-\infty }^{\infty }dt \int_0^{\infty }
dx f(t,x) \scl (t+x) \, , 
$$
where $f$ is a test function in $\R \times \R_+$. Introducing 
$$
{\widehat f}_\pm (k) \equiv \int_{-\infty }^{\infty }dt \int_0^{\infty }dx 
\, \e^{ik(t \pm x)}\, f(t,x) \quad , \neweq 
$$
by means of (A.2) and (2.15,16,30,31) one finds 
$$
\langle \scl (f)\Omega \, , \, \scl (f)\Omega \rangle 
= -2\int_0^{\infty } \mis \ln{k\over \mu_{{}_0} }\, {d\over dk} 
\left [\, \overline {{\widehat f}_+}(k) {\widehat f}_+(k)\, \right ] 
\quad , \neweq 
$$
$$
\langle \scr (f)\Omega \, , \, \scr (f)\Omega \rangle 
= -2\int_0^{\infty } \mis \ln{k\over \mu_{{}_0} }\, {d\over dk} 
\left [\,\overline {{\widehat f}_-}(k) {\widehat f}_-(k) \, \right ] 
\quad . \neweq 
$$
Varying $f$, the right hand sides of eqs.(2.44,45) have no definite sign, 
which shows that the metric is not positive definite. As in the case without 
boundary, this fact poses the problem of identifying a physical subspace 
with positive metric. We will face this fundamental problem in the next section. 
Since symmetries are essential for selecting the physical subspace, 
to the end of this section we will briefly analyze the symmetry 
content of the fields $\sc $ and $\dsc $ on the half line. 

As far as space-time symmetries are concerned, the situation is quite clear. 
Although dealing with a relativistic type dispersion relation, from 
the Poincar\'e group we are left on $\R \times \hl $ only with the time 
translations. Other admissible transformations, which form a group acting on 
$\R \times \hl $, are the dilatations 
$$
(t,x) \longmapsto (\varrho \, t , \varrho \, x ) \quad , 
\qquad \varrho > 0 \quad . 
\neweq 
$$
One can easily deduce from eqs.(2.35-37) that the rescaling (2.46) in any 
correlation function of the fields $\sc $ and $\dsc $ is equivalent to 
the substitutions $\mu \longmapsto \varrho \, \mu $ and 
$\eta \longmapsto \varrho \, \eta $.  

Let us investigate now the internal symmetries. From bosonization on 
the plane $\R \times \R$ we know that the shift operations 
$$
S\, :\, \sc (t,x) \longmapsto \sc (t,x) + \sigma \quad , \qquad 
\widetilde S\, :\, \dsc (t,x) \longmapsto \dsc (t,x) + \widetilde \sigma \quad , \qquad 
\sigma , \widetilde \sigma \in \R \quad , \neweq 
$$
are fundamental because they induce on fermion level   
the vector and axial transformations respectively. There exist {\it two nontrivial} 
charge operators $Q$ and $\widetilde Q$, which generate (2.47) and correspond to the 
conserved currents $\der_\nu \sc $ and $\der_\nu \dsc $ ($\nu = t,x$). 

The situation is quite different on $\R \times \hl $, because the boundary 
condition (2.4) is not invariant under both $S$ and $\widetilde S$. 
In fact, (2.4) is $S$-invariant only for $\eta = 0$, whereas for $0<\eta \leq \infty $ 
(2.4) is respected by $\widetilde S$ solely. 
As a consequence, for fixed $\eta $ one is left with {\it one nontrivial} 
charge. This essential new feature on the half line is confirmed by the 
explicit construction of $Q$ and $\widetilde Q$, we are going to describe now. 
The currents $\der_\nu \sc $ and $\der_\nu \dsc $ are conserved 
on $\R \times \hl $ as well, but the tentative to introduce $Q$ and 
$\widetilde Q$ in the conventional way by integrating $\der_t \sc $ and 
$\der_t \dsc $ on $\hl $ fails, since a boundary contribution at $x=0$ 
destroys the time-independence. One easily finds however the improved 
expressions 
$$
Q = \int_0^\infty dx \der_t \sc (t,x)  - \dsc (t,0) \quad , \qquad 
\widetilde Q = \int_0^\infty dx \der_t \dsc (t,x)  - \sc (t,0) 
\quad , \neweq 
$$
which do not have this drawback. In order to give rigorous meaning 
of eq.(2.48) on quantum level, we consider the finite volume charges  
$$
Q_r(t) = \int_0^\infty dx f_r(x) \der_t \sc (t,x)  - \dsc (t,0) = 
\int_0^\infty dx \left [\der_xf_r(x)\right ] \dsc (t,x)
\quad , \neweq 
$$
$$
{\widetilde Q}_r(t) = \int_0^\infty dx f_r(x) \der_t \dsc (t,x)  - \sc (t,0) 
= \int_0^\infty dx \left [\der_xf_r(x)\right ] \sc (t,x)
\quad , \neweq 
$$
where $f_r$ with $r > 0$ is any smooth function, such that  
$$
f_r(x) = \cases { 1 \, \, ,& $0 \leq x \leq r$ ; \cr 
0 \, \, , & $r\geq 2r$ . \cr } \neweq 
$$
Time-independence must be recovered in this framework in the limit 
$r \to \infty $. Indeed, a simple computation shows the existence of 
the following limits 
\medskip 
$$
\bigl [Q\, ,\, \sc (t,x) \bigr ] \equiv 
\lim_{r \to \infty} \bigl [Q_r(t_0 )\, ,\, \sc (t,x) \bigr ] 
= \cases { -i \, \, ,& $\eta = 0$ ; \cr 0 \, \, , & $0<\eta \leq \infty $ , \cr } 
\neweq 
$$ 
$$
\bigl [{\widetilde Q}\, ,\, \dsc (t,x) \bigr ] \equiv 
\lim_{r \to \infty} \bigl [{\widetilde Q}_r(t_0 )\, ,\, \dsc (t,x) \bigr ] 
= \cases { 0 \, \, ,& $\eta = 0$ ; \cr -i \, \, , & $0<\eta \leq \infty $ , \cr } 
\neweq 
$$
$$
\bigl [Q\, ,\, \dsc (t,x) \bigr ] \equiv 
\lim_{r \to \infty} \bigl [Q_r(t_0 )\, ,\, \dsc (t,x) \bigr ] = 0 
\quad , \neweq 
$$
$$
\bigl [{\widetilde Q}\, ,\, \sc (t,x) \bigr ] \equiv 
\lim_{r \to \infty} \bigl [{\widetilde Q}_r(t_0 )\, ,\, \sc (t,x) \bigr ] 
= 0 \quad , \neweq 
$$
which actually define the quantum charges $Q$ and $\widetilde Q$. 
Provided that $Q$ ($\widetilde Q$) is nonvanishing, it generates 
$S$ ($\widetilde S$). Moreover, in the range of $\eta $ where $Q$ is nontrivial, 
$\widetilde Q$ acts trivially and vice versa. This kind of complementarity 
suggests to introduce the charge $Q + \widetilde Q$, which defines 
according to eqs.(2.52-55), a nontrivial automorphism $\tau $ in the algebra 
generated by $\sc $ and $\dsc $. In the Fock representation $\F $, described 
above, the symmetry associated with $\tau $ is 
spontaneously broken. Indeed, from (2.52-55) one has 
$$
\langle \Omega \, ,\, \left [Q + 
{\widetilde Q}\, ,\, \sc (t,x) + \dsc (t,x) \right ]
\Omega \rangle = -i \not=  0 \quad , \neweq 
$$ 

In the absence of a boundary, the fields $\sc $ and $\dsc $ possess 
also a striking discrete symmetry under the exchange 
(duality) transformation $\sc (t,x) \longleftrightarrow \dsc (t,x)$. 
Equivalently, 
$$
\scr (t,x) \longmapsto \scr (t,x) \quad , \qquad 
\scl (t,x) \longmapsto - \scl (t,x) \quad , \neweq 
$$
which give the origin of the celebrated T-duality in string theory. Performing 
(2.57) in the two-point functions (2.35-37), we deduce that duality on the 
half line is broken for any fixed value of the boundary parameter $\eta $. 
We find in particular that the correlation functions for $\eta =0$ are related 
with those for $\eta = \infty $ by the transformation (2.57). In this sense 
the Neumann and Dirichlet boundary conditions are related by duality, which is 
in agreement also with eqs.(2.19,20). 

We conclude here our account on the basic features of the fields 
$\sc $ and $\dsc $ on\break $\R \times \hl $. In the next section we will 
apply these results for the construction of anyon fields in the 
presence of boundary.

\newchapt {Anyon Fields}

The concept of statistics in conventional quantum field theory is 
deeply related to locality and relativistic invariance. The latter 
is broken on $\R\times \hl $, but nevertheless, as shown in the 
previous section, the fields $\sc $ and $\dsc $ preserve quite 
remarkably some essential locality properties. We will see below that 
they are enough for constructing fields possessing in general anyonic 
statistics on the half line. Indeed, let us introduce for any 
couple $\zeta = (\alpha ,\beta )$ of real numbers the field 
$$
A(t,x;\zeta ) = 
z(\mu )\, :\exp [i\sqrt \pi (\alpha \sc + \beta \dsc )]: (t,x) 
\quad , \neweq 
$$
where the normal ordering $\, : \quad  :\, $ is taken with respect to 
the creation and annihilation operators 
$\{a^\ast (k), a(k) \, :\, k>0 \}$ and 
$$
z(\mu ) = \cases 
{ \mu^{\, \, {1\over 2}\alpha^2} \, \, ,& $\eta = 0$ ; \cr 
\mu^{\, \, {1\over 2}\beta^2} \, \, , & $0<\eta \leq \infty $ . \cr } 
\neweq 
$$
The normalization factor $z(\mu )$ is chosen for later convenience. 
Eq.(3.1) defines kind of ``vertex" operators on the half line, which 
are the main subject of this section. 

Let us concentrate first on the exchange properties of $A(t,x;\zeta )$. 
A standard calculation gives ($t_{12} \equiv t_1 -t_2$)
$$
A(t_1,x_1;\zeta_1 ) A(t_2,x_2;\zeta_2 ) = 
{\cal R}(t_{12},x_1,x_2;\zeta_1, \zeta_2 )\, 
A(t_2,x_2;\zeta_2 ) A(t_1,x_1;\zeta_1 ) 
\quad , \neweq 
$$
with exchange factor $\cal R$ given by 
$$
{\cal R}(t,x_1,x_2;\zeta_1, \zeta_2 ) = 
$$
$$
\exp \bigl \{ -i\pi \bigl [ 
(\alpha_1 \alpha_2 + \beta_1 \beta_2)D(t, x_1-x_2) + 
(\alpha_1 \alpha_2 - \beta_1 \beta_2)D_\eta (t, x_1+x_2) + 
$$
$$
(\alpha_1 \beta_2 + \alpha_2 \beta_1){\widetilde D}(t,x_1-x_2) + 
(\alpha_1 \beta_2 - \alpha_2 \beta_1){\widetilde D}_\eta (t,x_1+x_2)
\bigr ] \bigr \} \quad . \neweq 
$$
The statistics of the field (3.1) is governed by the behavior 
of (3.4) for $\zeta_1 = \zeta_2 = \zeta $ and space-like separated 
points $|t_{12}| < |x_1-x_2|$. 
Using eqs.(2.24-27) one gets in this domain 
$$
A(t_1,x_1;\zeta ) A(t_2,x_2;\zeta ) = 
\exp [-i\pi \alpha \beta \varepsilon (x_1-x_2)]
\, A(t_2,x_2;\zeta ) A(t_1,x_1;\zeta ) 
\quad . \neweq 
$$
Therefore $A(t,x;\zeta )$ is an anyon field whose statistics parameter is 
$$
\vartheta = \alpha \beta \quad . \neweq 
$$
One recovers Bose or Fermi statistics when $\vartheta $ is an even or odd 
integer respectively. The remaining values of $\vartheta $ lead to 
abelian braid statistics. The basic ingredient staying behind these 
generalized statistics is clearly the relative non-locality of 
$\sc $ and $\dsc $. 

Let us establish now the correlation functions of the $A$-fields, 
isolating, for further convenience, the $\mu $-dependence. The 
one-point function reads 
$$
\langle \Omega \, , \, A(t,x;\zeta ) \Omega \rangle = 1 \quad . \neweq 
$$ 
{}For $n\geq 2$ one finds 
$$
\langle \Omega \, , \, A(t_1,x_1;\zeta_1 )\cdots A(t_n,x_n;\zeta_n ) 
\Omega \rangle = \mu^{{1\over 2}s(\zeta_1,...,\zeta_n ;\eta )^2}  
\exp \sum_{i,j = 1\atop i<j}^n W_{ij}(t_{ij},x_i,x_j;\eta )
\quad , \neweq 
$$
where  
$$
s(\zeta_1,...,\zeta_n ;\eta ) = \cases{ 
\sum_{i=1}^n\alpha_i \, \, ,& $\eta = 0$ ; \cr 
\sum_{i=1}^n\beta_i \, \, , & $0< \eta \leq \infty $ . \cr } 
\neweq 
$$
The $\mu $-independent functions $W_{ij}$, given explicitly in 
Appendix B, are linear combinations 
of the distributions $u$ and $v_\pm $ (see eqs.(2.39,40)) with coefficients 
which are quadratic in the components of $\zeta_i$ and $\zeta_j$. From 
eqs.(3.8,9) one can extract relevant information about the 
existence of a physical subspace with positive metric. 
Indeed, let us select among all correlators (3.8), those which are
invariant under the automorphism $\tau $ generated by $Q + {\widetilde Q}$. 
One easily finds that they are characterized by the condition 
$$
s(\zeta_1,...,\zeta_n ;\eta ) = 0 \quad , \neweq 
$$
which implies that the $\tau $-invariant vacuum expectation values of the 
$A$-fields do not depend on the infra-red parameter $\mu $.
Since negative metric states have infra-red origin, one may expect that
$\tau $-invariance leads to positive definiteness. This last property is 
crucial for the physical interpretation and needs a careful treatment. 
The direct investigation of the functions (3.8) is a hard job, because 
$W_{ij}$ are quite involved (see eqs.(B.1-3)). For this 
reason we will adapt to our case the approach developed in [13,23] for 
studying positivity in the massless Thirring model on the plane. 

Let us consider the algebra $\palg $ generated by 
$\{a^\ast_\lambda (k), a_\lambda (k)\, :\, k\geq 0 \}$, 
satisfying the commutation relations 
$$
[a_\lambda (k)\, ,\, a_\lambda (p)] = 
[a^\ast_\lambda (k)\, ,\, a^\ast_\lambda (p)] = 0 
\quad , \neweq 
$$
$$ 
[a_\lambda (k)\, ,\, a^\ast_\lambda (p)] = 
2\pi \theta (k) (k + \lambda )^{-1}\, \delta (k-p)  
\quad , \neweq 
$$
where $\lambda > 0$ has the dimension of a mass. Since 
$\theta (k) (k + \lambda )^{-1}$ is a positive definite distribution, 
$\palg $ admits a positive metric Fock representation $\F_\lambda $. 
We denote by $(\cdot \, ,\, \cdot )_{{}_\lambda }$ and $\Omega_\lambda $ the 
scalar product and vacuum state in $\F_\lambda $ respectively and define 
$\psc $ and $\pdsc $ by substituting $\{a(k) , a^\ast (k)\}$ in eqs.(2.17,18) 
with $\{a_\lambda (k) , a^\ast_\lambda  (k)\}$. Compared to 
$\sc $ and $\dsc $, the fields $\psc $ and $\pdsc $ still satisfy eqs.(2.1,3), 
but have worse locality properties. Nevertheless, they are very 
useful for facing the positivity problem. For this purpose we introduce 
$$
A_\lambda (t,x;\zeta ) = 
z\left (\lambda \e^{\gamma_{{}_E}} \right )
\, :\exp [i\sqrt \pi (\alpha \psc + \beta \pdsc )]: (t,x) 
\quad , \neweq 
$$
where the normal ordering $\, :\quad :\, $ refers to 
$\{a_\lambda (k) , a^\ast_\lambda (k)\}$. The main idea now 
is to relate the expectation values of the $A_\lambda $-fields 
with those of $A$ and use the positivity of the metric in $\F_\lambda $. 
In doing that we first need the $\lambda $-counterparts 
of the correlation functions (2.35-40). One easily finds 
$$
w^\lambda_{{}_{RR}}(\xi_1, \xi_2) = w^\lambda_{{}_{LL}}(\xi_1, \xi_2) = 
v(\lambda \xi_{12}) \quad , \neweq 
$$
$$
w^\lambda_{{}_{RL}}(\xi_1, \xi_2) = \cases 
{v(\lambda \xi_{12})\, \, , &$\eta = 0$;\cr 
{\lambda + i\eta \over \lambda - i\eta }\, v(\lambda \xi_{12}) + 
{i\eta \over \lambda - i\eta }\, v_-(\eta \xi_{12}) \, \, , 
&$0<\eta < \infty $;\cr - v(\lambda \xi_{12}) 
&$\eta = \infty $,\cr } \neweq  
$$
$$
w^\lambda_{{}_{LR}}(\xi_1, \xi_2) = \cases 
{v(\lambda \xi_{12})\, \, , &$\eta = 0$;\cr 
{\lambda - i\eta \over \lambda + i\eta }\, v(\lambda \xi_{12}) - 
{i\eta \over \lambda + i\eta }\, v_+(-\eta \xi_{12}) \, \, , 
&$0<\eta < \infty $;\cr - v(\lambda \xi_{12}) 
&$\eta = \infty $.\cr } \neweq  
$$
The new function $v$, appearing in (3.14-16), is 
$$
v(\xi ) = -{1\over \pi }\, \e^{i\xi }
\, {\rm Ei} (-i\xi - \epsilon ) 
\quad . \neweq 
$$
By means of eqs.(3.14-17) it is not difficult to show at this point that 
$$ 
\langle \Omega \, , \, A(t_1,x_1;\zeta_1 )\cdots A(t_n,x_n;\zeta_n ) \Omega \rangle
= \lim_{\lambda \to 0}\,  
(\Omega_\lambda \, , \, A_\lambda (t_1,x_1;\zeta_1 )\cdots A_\lambda (t_n,x_n;\zeta_n ) 
\Omega_\lambda )_{{}_\lambda } \quad , \neweq 
$$
{\it provided that} (3.10) {\it holds}. The limit in eq.(3.18) is in the sense of 
distributions. Since the expectation values of $A_\lambda $ are positive definite by 
construction, from (3.18) we infer that the same is true for those correlators 
(3.8), which are $\tau $-invariant. This concludes the argument. 

Besides ensuring positivity, in the cases of Neumann and Dirichlet boundary 
conditions, the $\tau $-invariance implies in addition  
the following factorization property under dilatations (2.46). Indeed, 
assuming that $s(\zeta_1,...\zeta_{k+m} ;\eta ) = 0$, one finds  
$$
\lim_{\varrho  \downarrow 0} 
\varrho ^{{1\over 2} q(\zeta_{k+1},...,\zeta_{k+m}; \eta )}  
\langle \Omega\, ,\, A(t_1,x_1;\zeta_1)\cdots A(t_k,x_k;\zeta_k ) \, \cdot 
$$
$$
A(\varrho t_{k+1}, \varrho x_{k+1} ; \zeta_{k+1}) \cdots  
A(\varrho t_{k+m}, \varrho x_{k+m} ; \zeta_{k+m}) \Omega \rangle 
$$
$$
=\cases{ 
\langle \Omega , A(t_1,x_1;\zeta_1) \cdots A(t_k,x_k;\zeta_k) 
\Omega \rangle \, \cdot & \cr
\langle \Omega , A(t_{k+1},x_{k+1};\zeta_{k+1})\cdots  
A(t_{k+m},x_{k+m};\zeta_{k+m})\Omega \rangle  \, , 
& if $\, s(\zeta_1 ,...,\zeta_k ;\eta ) = 0 $  ; \cr 
0 \, , & if $\, s(\zeta_1 ,...,\zeta_k ;\eta ) \not= 0 $ , \cr}
\neweq 
$$
where
$$
q(\zeta_1,...,\zeta_n ;\eta ) = \cases{ 
\sum_{i=1}^n\alpha_i^2 \, \, ,& $\eta = 0$ ; \cr 
\sum_{i=1}^n\beta_i^2 \, \, , & $\eta = \infty $ . \cr } 
\neweq 
$$ 
We will discuss the physical relevance of eq.(3.19) few lines below. 

The results of this section allow to construct a physical (positive metric) 
representations of the anyon fields under consideration. One can proceed 
as follows. Fixing a set $\cZ = \{\zeta_1, -\zeta_1,...,\zeta_n, -\zeta_n \}$, 
we consider the family of all $\tau $-invariant 
correlation functions of $\{ A(t,x;\zeta )\, :\, \zeta \in \cZ\}$. 
Via the reconstruction theorem [20], this family determines (up to 
unitary equivalence) a Hilbert space $\{\H_{\rm ph} , \, (\cdot \, ,\, \cdot ) \}$, 
a vacuum state $\Omega_{\rm ph} \in \H_{\rm ph} $ and quantum fields 
$\{A_{\rm ph}(t,x;\zeta )\, :\, \zeta \in \cZ \}$, such that 
$$ 
\left (\Omega_{\rm ph} \, ,\, A_{\rm ph}(t_1,x_1;\zeta_{i_1}) 
\cdots A_{\rm ph}(t_k,x_k;\zeta_{i_k}) \Omega_{\rm ph} \right ) 
$$
$$
=\cases{ 
\langle \Omega , A(t_1,x_1;\zeta_{i_1}) \cdots A(t_k,x_k;\zeta_{i_k}) 
\Omega \rangle 
 \, ,\, & if $\, s(\zeta_{i_1} ,...,\zeta_{i_k} ;\eta ) = 0 $  ; \cr 
0 \, ,\, & if $\, s(\zeta_{i_1} ,...,\zeta_{i_k} ;\eta ) \not= 0 $ . \cr}
\neweq
$$
The vacuum expectation values (3.21) are invariant under time translations 
and admit an analytic continuation in the domain defined by shifting any time variable 
according to $t_j \mapsto \tau_j -i\sigma_j $ with 
$\sigma_1 > \sigma_2 > ... > \sigma_k$. 
The correlation functions, obtained in this way, satisfy a kind of cluster 
property under translation of $\sigma $-clusters, respecting 
the order of $\sigma_j$. This fact suggests [20] that the physical 
representation at hand is irreducible. 

In the special cases $\eta =0$ and $\eta = \infty $, the correlation functions 
(3.21) do not involve dimensional parameters and one can easily check that the dilatations 
(2.46) are unitarily implemented. In other words, there exists a unitary 
representation $U(\varrho )$ of the dilatation group in $\H_{\rm ph}$, such that: 
$$
U(\varrho ) \Omega_{\rm ph} = \Omega_{\rm ph} \quad , \neweq 
$$
$$
U(\varrho )\, A_{\rm ph}(t,x;\zeta )\, U(\varrho )^{-1} = 
\varrho^{d_\zeta }\, A_{\rm ph}(\varrho t,\varrho x;\zeta ) \quad , 
\qquad \zeta \in \cZ \quad , 
\neweq 
$$
where $d_\zeta $ is the scale dimension given by 
$$
d_\zeta  = \cases { {1\over 2}\alpha^2\, \, , & $\eta = 0$; \cr 
{1\over 2}\beta^2\, \, , & $\eta = \infty $. \cr } \neweq 
$$
Combining eq.(3.19) and eqs.(3.21-24), one gets also that 
$$
\lim_{\varrho  \downarrow 0} \, 
(\Omega_{\rm ph}\, ,\, A_{\rm ph}(t_1,x_1;\zeta_{i_1}) 
\cdots A_{\rm ph}(t_k,x_k;\zeta_{i_k}) \, U(\varrho) \cdot 
$$
$$ 
A_{\rm ph}(t_{k+1}, x_{k+1};\zeta_{i_{k+1}}) \cdots  
A_{\rm ph}(t_{k+m}, x_{k+m};\zeta_{i_{k+m}}) \Omega_{\rm ph} ) 
$$
$$
= (\Omega_{\rm ph} , A_{\rm ph}(t_1,x_1;\zeta_{i_1}) \cdots A_{\rm ph}(t_k,x_k;\zeta_{i_k}) 
\Omega_{\rm ph} )\, \cdot 
$$
$$ 
(\Omega_{\rm ph} , A_{\rm ph}(t_{k+1},x_{k+1};\zeta_{i_{k+1}})\cdots 
A_{\rm ph}(t_{k+m},x_{k+m};\zeta_{i_{k+m}})\Omega_{\rm ph} ) \quad . 
\neweq
$$ 
Eq.(3.25) represents a sort of cluster decomposition property 
related to dilatations instead to the conventionally considered [20] 
translations in space. Accordingly, a straightforward modification of 
the argument in Theorem 3-7 of [20] implies that $\Omega_{\rm ph}$ is the 
unique dilatation invariant state in $\H_{\rm ph}$. 

Summarizing, given a set $\cZ$, we have constructed a physical (positive 
metric) representation of the associated anyon fields. In the sequel we will 
use such representations for bosonization and more generally, for studying 
quantum field theory models on the half line. For conciseness we shall 
omit the index ``ph", but one should keep in mind that $A$ and $A_{\rm ph}$ 
are different fields. Finally, 
in what follows we will have the occasion to use anyon fields simultaneously 
involving $\sc_1 $ and $\sc_2 $ associated with two 
different boundary parameters $\eta_1 $ and $\eta_2 $. 
Let us consider for instance 
$$
A^\prime (t,x;\zeta ) = 
\mu^{{1\over 2}(\alpha^2 + \beta^2)} 
\, :\exp \{ i\sqrt \pi [\alpha (\sc_1 + \sc_2) + \beta (\dsc_1 + \dsc_2 )]\}: (t,x) 
\quad , \neweq 
$$
with $\eta_1 = 0$ and $\eta_2 = \infty $. In this case 
the correlation functions 
$$
\langle \Omega \, , \, A^\prime (t_1,x_1;\zeta_1 )\cdots A^\prime (t_n,x_n;\zeta_n ) 
\Omega \rangle 
$$ 
are $\tau $-invariant only if both selection rules 
$$
\sum_{i=1}^n\alpha_i = 0 \quad , \qquad  
\sum_{i=1}^n\beta_i = 0 \quad , \neweq 
$$
are satisfied. Taking into account (3.27), the construction of the 
physical representation of 
$\{A^\prime (t,x;\zeta )\, :\, \zeta \in \cZ \}$ follows the scheme 
developed above.

\newchapt {Bosonization}

In this section we address the problem of bosonization of the free 
massless Dirac field $\psi $ on the half line $\hl $. 
The equation of motion is 
$$
(\gamma_t \der_t - \gamma_x \der_x)\psi (t,x) = 0 \quad , \neweq 
$$
where
$$
\psi (t,x)=\pmatrix{ \psi_1(t,x) \cr \psi_2(t,x) \cr} 
\quad, \qquad 
\gamma_t=\pmatrix{ 0 & 1 \cr 1 & 0 \cr} \quad, \qquad
\gamma_x=\pmatrix{ 0 & 1 \cr -1 & 0 \cr} \quad .\neweq
$$
The standard vector and axial currents are 
$$
j_\nu (t,x) = \overline \psi (t,x) \gamma_\nu \psi (t,x) \quad , 
\qquad j_\nu^5 (t,x) = \overline \psi (t,x) \gamma_\nu \gamma^5 \psi (t,x) \quad , 
\qquad \nu = t,x \quad , \neweq 
$$
with $\psi \equiv \psi^\ast \gamma_t $ and $\gamma^5 \equiv -\gamma_t\gamma_x $. 
{}From eq.(4.1) it follows that both $j_\nu $ and $j_\nu^5$ are conserved. 
Moreover, the $\gamma^5$-identities $\gamma_t\gamma^5 = - \gamma_x $ and 
$\gamma_x \gamma^5 = - \gamma_t $ imply the relations 
$$
j_t^5 = - j_x \quad , \qquad j_x^5 = -j_t \quad . 
\neweq 
$$
 
Our main goal below will be to quantize (4.1) in terms of 
the fields $\sc $ and $\dsc $, establishing the 
boundary conditions on $\psi $ encoded in 
the parameter $\eta $. For this purpose we set 
$$
\psi_1 (t,x) = \chi_{{}_1} A(t,x;\zeta_1 = (\alpha ,\alpha )) = 
\chi_{{}_1} z(\mu ) :\exp(i\sqrt \pi \alpha \scr ): (t-x)   
\quad , \neweq 
$$
$$
\psi_2 (t,x) = \chi_{{}_2} A(t,x;\zeta_2 = (\alpha ,-\alpha )) = 
\chi_{{}_2} z(\mu ) :\exp(i\sqrt \pi \alpha \scl ): (t+x) 
\quad , \neweq 
$$
where $\chi_{{}_1},\, \chi_{{}_2} \in \C$ and $0\leq \eta \leq \infty$ are for 
the moment free parameters. One easily 
shows that $\psi $, given by eqs.(4.5,6), satisfies the Dirac equation. 
As a direct consequence of eqs.(3.3,4), one also finds that 
$\psi $ has Fermi statistics  
$$
\psi_i (t_1,x_1) \psi_j (t_2,x_2) = - \psi_j (t_2,x_2) \psi_i (t_1,x_1) 
\quad , \qquad |t_{12}| < |x_1 - x_2| \quad , \neweq 
$$
provided that 
$$
\alpha^2 = 2k+1 \quad , \qquad k \in \N \quad . \neweq 
$$
We impose the condition (4.8) throughout this section. 

The next step is to define the quantized currents (4.3). We will 
use the point splitting procedure [24], starting from the expression 
$$
j_\nu (t,x) = {1\over 2} \lim_{\epsilon_x \downarrow 0} Z(\epsilon_x ;\alpha ) 
\left [\, \overline \psi (t,x)\gamma_\nu \psi (t, x+\epsilon_x ) + 
\overline \psi (t, x+\epsilon_x ) \gamma_\nu \psi (t,x)\, \right ] 
\quad , \neweq 
$$
where $Z(\epsilon_x ;\alpha )$ implements the renormalization. 
Separating only the space coordinate $x$ in (4.9) 
turns out to be enough and is quite natural on $\R \times \hl$, 
where Lorentz invariance is any way broken. The basic general formula for 
evaluating (4.9), is obtained by normal ordering the product 
$A^\ast (t,x+\epsilon_x;\zeta ) A(t,x;\zeta )$. One has 
$$
A^\ast (t,x+\epsilon_x;\zeta ) A(t,x;\zeta ) = 
$$
$$ 
z(\mu )^2 
:\exp \Bigl \{ i\sqrt \pi \bigl [ \alpha \sc (t,x) - \alpha \sc (t,x+\epsilon_x ) 
+ \beta \dsc (t,x) - \beta \dsc (t,x+\epsilon_x ) \bigr ] \Bigr \}: 
$$
$$
\exp \Bigl \{ {\pi \over 4} \bigl [ (\alpha + \beta )^2 w_{{}_{RR}} (-\epsilon_x ) 
+ (\alpha - \beta )^2 w_{{}_{LL}} (\epsilon_x ) 
+ (\alpha^2 - \beta^2 )\bigl ( w_{{}_{RL}} (-2x - \epsilon_x )
+ w_{{}_{LR}}(2x + \epsilon_x )\bigr ) \bigr ] \Bigr \}\, . 
\neweq 
$$
Eq.(4..10) keeps manifestly trace of the new intrinsic features of the theory on 
the half line, namely, the breakdown of translation invariance and the lack 
of left-right factorization, detected in section 2. At this point one should take 
into account however, that for the particular values of $\zeta_1$ and $\zeta_2$ 
in eqs.(4.5,6), the $w_{{}_{LR}}$ and $w_{{}_{RL}}$ contributions in (4.10) drop out. 
Assuming the normalization 
$$
|\chi_{{}_1}|^2 = |\chi_{{}_2}|^2 = {1\over 2\pi } \quad , \neweq 
$$ 
and setting 
$$
Z(\epsilon_x ;\alpha ) = 
{\epsilon_x^{\alpha^2 -1}\over \alpha \sin \left ({\pi\over 2}\alpha^2\right ) }
\quad , \neweq 
$$
one finds the conserved current 
$$
j_\nu (t,x) = - {1\over \sqrt \pi}\der_\nu \sc (t,x) \quad , \neweq 
$$
thus recovering the same type of relation as in the conventional 
bosonization on the whole line [18,19]. 

In analogy with (4.9) we introduce the axial current by  
$$
j_\nu^5 (t,x) = {1\over 2} \lim_{\epsilon_x \downarrow 0} Z(\epsilon_x ;\alpha ) 
\left [\, \overline \psi (t,x)\gamma_\nu \gamma^5 \psi (t, x+\epsilon_x ) + 
\overline \psi (t, x+\epsilon_x )\gamma_\nu \gamma^5 \psi (t,x)\, \right ]  
\quad . \neweq 
$$ 
The vector current result and the $\gamma^5$-identities directly imply that 
the limit in the right hand side of (4.14) exists and 
$$
j^5_\nu (t,x) = - {1\over \sqrt \pi}\der_\nu \dsc (t,x) \quad . \neweq 
$$ 
The classical relations (4.4) are therefore respected also on quantum level. 

Eqs.(4.13,15) have various important consequences. First of all they 
imply that the boundary conditions on $\psi $ are most 
conveniently formulated in terms of the currents. Indeed, 
according to eqs.(2.19,20) one has that 
$$
\lim_{x \downarrow 0}j_x (t,x) = 0 \neweq 
$$
is realized when $\eta = 0$, whereas 
$$
\lim_{x \downarrow 0}j_x^5 (t,x) = 0 \neweq 
$$
holds for $\eta = \infty $. In the domain $0< \eta < \infty $ one has 
instead 
$$
\lim_{x \downarrow 0}(\der_x - \eta ) j_x^5 (t,x) = 0 \quad . \neweq 
$$
In verifying (4.18), one has to exchange the order of two operations - 
a limit and a derivative. By inspection, this is correct in the case 
under consideration.  

Using our previous results on the charges (2.52-55), we can investigate now 
the impact of the boundary conditions (4.16-18) on the symmetry content 
of our bosonization construction. We will concentrate for simplicity 
on the values $\eta = 0$ and $\eta = \infty $. Combining eqs.(2.4,19,20) 
and (2.49,50), in these cases the charges 
$$
\Q = -{1\over \alpha \sqrt \pi} Q \quad , \qquad 
\Q^5 = -{1\over \alpha \sqrt \pi} \widetilde Q \quad , \neweq 
$$
can be expressed entirely in terms of $j_t $ or $j_t^5$. 

Let us examine first the $\eta = 0$ (vector) phase. One finds 
$$ 
\left [\Q \, ,\, \psi (t,x) \right ] = 
- \psi (t,x) 
\quad , \qquad 
\left [\Q \, ,\, \overline \psi (t,x) \right ] = 
\, \, \overline \psi (t,x) 
\quad , \neweq 
$$
which induce the conventional automorphism associated with the vector 
current. Following the general scheme developed in the previous section, 
the physical representation is reconstructed from the 
$\Q$-invariant correlation functions of $\psi $ and $\overline \psi $. 
According to eqs.(2.53,55), the chiral charge $\Q^5$ 
generates in this phase the identity automorphism, which allows for 
a left-right mixing. For example, the only nontrivial two-point functions are 
$$
\langle \Omega \, ,\, \psi_i^* (t_1,x_1)\, \psi_j (t_2,x_2)\Omega \rangle = 
{1\over 2\pi }
\pmatrix{ {1\over [it_{12} - i(x_1-x_2) + \epsilon ]^{\alpha^2 }} & 
{2\pi \overline \chi_{{}_1} \chi_{{}_2} \over [it_{12} - i(x_1+x_2) + \epsilon ]^{\alpha^2 }} \cr 
{2\pi \chi_{{}_1} \overline \chi_{{}_2} \over [it_{12} + i(x_1+x_2) +\epsilon ]^{\alpha^2 }} & 
{1\over [it_{12} + i(x_1-x_2) + \epsilon ]^{\alpha^2 }}\cr} \quad , 
\neweq 
$$
\medskip 
$$
\langle \Omega \, ,\, \psi_i (t_1,x_1)\, \psi_j^* (t_2,x_2)\Omega \rangle = 
{1\over 2\pi }
\pmatrix{ {1\over [i t_{12} - i(x_1-x_2) + \epsilon ]^{\alpha^2 }} & 
{2\pi \chi_{{}_1} \overline \chi_{{}_2} \over [it_{12} - i(x_1+x_2) +\epsilon ]^{\alpha^2 }} \cr 
{2\pi \overline \chi_{{}_1} \chi_{{}_2} \over [it_{12} + i(x_1+x_2) +\epsilon ]^{\alpha^2 }} & 
{1\over [it_{12} + i(x_1-x_2) +\epsilon ]^{\alpha^2 }}\cr} \quad . 
\neweq 
$$
\medskip 
\noindent The off-diagonal elements of (4.21,22) describe the left-right mixing. 
The mixing parameter ${\overline \chi_{{}_1} \chi_{{}_2}}$
is clearly invariant under vector transformations.

{}For $\eta = \infty $ (axial phase) one has in some sense 
the opposite situation: the charge $\Q$ acts trivially, 
but $\Q^5$ generates the axial transformations 
$$
\left [\Q^5\, ,\, \spr (t,x) \right ] = 
- \spr (t_2,y) 
\quad , \qquad 
\left [\Q^5\, ,\, \spl (t,x) \right ] = 
\, \, \spl (t,x) 
\quad , \neweq 
$$
where 
$$
\spr \equiv {1\over 2}(1 + \gamma^5) \psi = \psi_1 \quad , \qquad 
\spl \equiv {1\over 2}(1 - \gamma^5) \psi = \psi_2 \quad . \neweq 
$$
The $\Q^5$-invariant correlators give rise to the physical representation. 
In this phase the 
left-right mixing is forbidden, but there exist nontrivial correlators 
with nonvanishing total $\Q$-charge. The complete set of two-point functions is: 
\medskip 
$$
\langle \Omega \, ,\, \psi_i (t_1,x_1) \, \psi_j (t_2,x_2) \Omega \rangle =
{\chi_{{}_1}\chi_{{}_2}} \pmatrix {0 &
{1 \over [it_{12}-i(x_1+x_2) +\epsilon]^{\alpha^2}} \cr
{1 \over [it_{12}+i(x_1+x_2) +\epsilon ]^{\alpha^2}}& 0 \cr}
\quad , \neweq 
$$ 
$$
\langle \Omega \, ,\, \psi_i^* (t_1,x_1) \, \psi_j^* (t_2,x_2) \Omega \rangle =
{\overline \chi_{{}_1} \overline \chi_{{2}}} \pmatrix {0 &
{1 \over [it_{12}-i(x_1+x_2) +\epsilon]^{\alpha^2}} \cr
{1 \over [it_{12}+i(x_1+x_2) +\epsilon ]^{\alpha^2}}& 0 \cr}
\quad , \neweq 
$$  
$$
\langle \Omega \, ,\, \psi_i (t_1,x_1) \, \psi_j^*(t_2,x_2) \Omega \rangle =
\langle \Omega \, ,\, \psi_i^* (t_1,x_1) \, \psi_j (t_2,x_2) \Omega \rangle =
$$  
$$
={1\over 2\pi } \pmatrix {
{1\over [it_{12}-i(x_1-x_2) +\epsilon ]^{\alpha^2}} & 0 \cr 0 &
{1\over [it_{12}+i(x_1-x_2) +\epsilon ]^{\alpha^2}} \cr} \quad . \neweq 
$$
\medskip 
\noindent The mixing parameter is now ${\chi_{{}_1} \chi_{{}_2}}$ and is 
invariant under axial transformations. 

This is the right point to clarify the role of the duality in the above 
scheme. Translating the transformation (2.57) in terms of $\psi $, 
one finds 
$$
\psi_1 (t,x) \longmapsto \psi_1 (t,x) \quad , \qquad 
\psi_2 (t,x) \longmapsto {\chi_{{}_2} \over \overline \chi_{{}_2} }\, \psi_2^* (t,x) 
\quad . \neweq 
$$
Using (4.28), one can verify that duality 
maps the vector phase in the axial one and vice versa. The two-point functions 
(4.21,22,25-27) provide useful checks. 

In the light of these results, one may ask if fields of the type (3.26), 
combining different boundary conditions, can provide a 
bosonized solution of eq.(4.1) in which {\it both} vector and chiral 
symmetry are nontrivially implemented (vector-axial phase). For answering 
this question we consider 
$$
\psi_1 (t,x) =  {z(\mu) \over \sqrt{2\pi} } 
:\exp \left [i\alpha \sqrt {\pi \over 2} ( \sc_{1_R} + \sc_{2_R} )\right ]: (t-x)   
\quad , \neweq 
$$
$$
\psi_2 (t,x) =   {z(\mu) \over \sqrt{2\pi} }
:\exp \left [i\alpha \sqrt {\pi \over 2} ( \sc_{1_L} + \sc_{2_L} )\right ]:(t+x) 
\quad , \neweq 
$$ 
where $\sc_1$ and $\sc_2$ satisfy Neumann and Dirichlet boundary 
conditions respectively. 
It is easily seen that eqs.(4.29,30) define a solution of eq.(4.1) 
with Fermi statistics. 
Defining the currents as in (4.9,14), with the renormalization constant
given by (4.12), we obtain 
$$
j_\nu (t,x) = - {1\over \sqrt {2\pi}}\left [\der_\nu \sc_1 (t,x) + 
\der_\nu \sc_2 (t,x) \right ] 
\quad , \neweq 
$$
$$
j_\nu^5 (t,x) = - {1\over \sqrt {2\pi}}\left [\der_\nu \dsc_1 (t,x) + 
\der_\nu \dsc_2 (t,x) \right ] 
\quad . \neweq 
$$
The corresponding charges generate (4.20) and (4.23) 
simultaneously, which answers affirmatively our question. 
The correlation functions defining the physical representation 
are both $\Q$ and $\Q^5$-invariant. For instance, the only nontrivial 
two-point functions are (4.27). 

In order to find the relative boundary condition, we proceed as follows.  
The\break solution (4.29,30) is based on two copies of creation 
and annihilation operators\break $\{a_1^*(k),\, a_1(k)\, : \, k>0\}$ and 
$\{a_2^*(k),\, a_2(k)\, : \, k>0\}$, satisfying (2.15,16) and 
acting in the Fock spaces $\F_1 $ and $\F_2 $ respectively. Since 
these Fock representations are unitary equivalent, there exists 
a unitary operator $U \, :\, \F_1 \rightarrow \F_2 $, such that 
$$
a_1(k) = U\, a_2(k)\, U^{-1} \quad , \qquad 
a_1^*(k) = U\, a_2^*(k)\, U^{-1} \quad . \neweq 
$$
Using this fact and eqs.(2.19,20), one concludes that the solution 
(4.29,30) satisfies 
$$
\lim_{x \downarrow 0}j_x (t,x) = 
U\left [\lim_{x \downarrow 0}j^5_x (t,x) \right ]U^{-1} 
\quad . \neweq 
$$
which represents a sort of vector-axial symmetric boundary condition.  

It is worth mentioning that dilatations give rise to an exact symmetry 
in any of the three phases considered above and 
$$
d_\psi = {1\over 2} \alpha^2 = k + {1\over 2} \quad , 
\qquad k \in \N \quad . \neweq 
$$

Summarizing, we established a bosonization procedure for 
the free massless Dirac equation 
in $\R \times \hl $. An essential new feature 
is the close relationship between boundary conditions 
on the scalar potential $\sc $ and the symmetry content of the corresponding 
physical representation of $\psi $. Taking as an example the massless 
Thirring model, we will discuss in the next section the new features 
related to the bosonization of fermion fields satisfying 
nonlinear equations on the half line.

\newchapt {The Thirring Model on the Half Line}

The classical equation of motion for the massless Thirring model [25] 
is 
$$
i(\gamma_t \der_t - \gamma_x \der_x)\Psi (t,x) = 
g\left [\gamma_t J_t(t,x) - 
\gamma_x J_x (t,x) \right ] \Psi (t,x) 
\quad , \neweq 
$$
where $g\in \R $ is the coupling constant and $J_\nu$ is 
the conserved current 
$$
J_\nu (t,x) = \overline \Psi (t,x) \gamma_\nu \Psi (t,x) \quad . 
\neweq 
$$
As well known, the axial current 
$$
J_\nu^5 (t,x) = \overline \Psi (t,x) \gamma_\nu \gamma^5 \Psi (t,x)  
\neweq 
$$
is also conserved. We shall investigate below the Thirring model 
on $\R \times \hl $, concentrating first to the phases corresponding to 
the {\it vector} 
$$
\lim_{x \downarrow 0}J_x (t,x) = 0 \quad , \neweq 
$$
and the {\it axial} boundary condition
$$
\lim_{x \downarrow 0}J^5_x (t,x) = 0 \neweq 
$$
respectively. 

{}For quantizing the system, we consider the fields 
$$
\Psi_1 (t,x) = \chi_{{}_1} A(t,x;\zeta_1 = (\alpha,\beta )) = 
\chi_{{}_1} z(\mu ) :\exp[i\sqrt \pi (\alpha \sc + \beta \dsc )]: (t,x)   
\quad , \neweq 
$$
$$
\Psi_2 (t,x) = \chi_{{}_2} A(t,x;\zeta_2 = (\alpha ,-\beta )) = 
\chi_{{}_2} z(\mu ) :\exp[i\sqrt \pi (\alpha \sc - \beta \dsc )]: (t,x)  
\quad . \neweq 
$$ 
The main point is the construction of the quantum currents 
$J_\nu $ and $J_\nu^5 $. In analogy with (4.9) we set 
$$
J_\nu (t,x) = {1\over 2} \lim_{\epsilon_x \downarrow 0} 
Z_\nu (x, \epsilon_x ;\alpha ,\beta )
(\left [\, \overline \Psi (t,x)\gamma_\nu \Psi (t, x+\epsilon_x ) + 
\overline \Psi (t, x+\epsilon_x ) \gamma_\nu \Psi (t,x)\, \right ] 
\quad . \neweq 
$$
There is no summation over $\nu $ in (5.8) and the presence of two $x$-dependent 
renormalization constants $Z_t$ and $Z_x$ reflects the fact both 
translation and Lorentz invariance are broken. We also define 
$$
J_\nu^5 (t,x) = {1\over 2} \lim_{\epsilon_x \downarrow 0} 
Z^5_\nu (x, \epsilon_x ;\alpha ,\beta )
(\left [\, \overline \Psi (t,x)\gamma_\nu \gamma^5 \Psi (t, x+\epsilon_x ) + 
\overline \Psi (t, x+\epsilon_x ) \gamma_\nu \gamma^5 \Psi (t,x)\, \right ] 
\quad , \neweq 
$$
$Z_\nu^5$ being the chiral counterpart of $Z_\nu $. 
Our aim now will be to determine the parameters $\eta \geq 0$ and  
$\alpha ,\beta \in \R$, such that: 

\item {(i)} $\Psi $ obeys Fermi statistics; 

\item {(ii)} there exist suitable renormalization functions $Z_\nu$ and 
$Z_\nu^5$ for which the limits in the right hand sides of 
eqs.(5.8,9) exist and give 
$$
J_\nu (t,x) = {-1\over \sqrt \pi } \der_\nu \sc (t,x) \quad , \qquad 
J_\nu^5 (t,x) = {-1\over \sqrt \pi } \der_\nu \dsc (t,x) 
\quad ; \neweq 
$$

\item {(iii)} the quantum equation of motion, which in agreement with (5.10) 
takes the form 
$$
i(\gamma_t \der_t - \gamma_x \der_x)\Psi (t,x) = 
-{g\over \sqrt \pi } : \left (\gamma_t \der_t \sc - 
\gamma_x \der_x \sc \right ) \Psi : (t,x)  \quad , \neweq 
$$
is satisfied; 

\item {(iv)} either the vector or the axial boundary condition is fulfilled. 

\noindent {}From (i) one can easily deduce that  
$$
\alpha \beta = 2k + 1 \quad , \qquad k \in \Z \quad . \neweq 
$$
Provided that (ii) holds, point (iv) implies 
$$
\eta =  \cases 
{0\, \, , & vector b.c. \cr 
\infty \, \, , & axial b.c. \cr } 
\neweq  
$$ 
{}For these values of $\eta $, the general formula (4.10) gives 
$$
{1\over 2}\left [\Psi_1^*(t,x+\epsilon_x ) \Psi_1 (t,x) + 
\Psi_1^*(t,x) \Psi_1 (t,x+\epsilon_x ) \right ] = 
$$
$$
{1\over 2\sqrt \pi } K(x, \epsilon_x ;\alpha ,\beta ) 
\left \{ \epsilon_x 
\left [\alpha \der_x \sc (t,x) - \beta \der_t \sc (t,x)\right ] 
+ O(\epsilon_x^2 ) \right \} \quad , \neweq 
$$
$$
{1\over 2}\left [\Psi_2^*(t,x+\epsilon_x ) \Psi_2 (t,x) + 
\Psi_2^*(t,x) \Psi_2 (t,x+\epsilon_x ) \right ] = 
$$
$$
{1\over 2\sqrt \pi } K(x, \epsilon_x ;\alpha ,\beta ) 
\left \{ - \epsilon_x 
\left [\alpha \der_x \sc (t,x) + \beta \der_t \sc (t,x) \right ] 
+ O(\epsilon_x^2 ) \right \} \quad , \neweq 
$$
where 
$$
K(x, \epsilon_x ;\alpha ,\beta ) = \cases {  
\sin \left ({\pi \over 2}\alpha \beta \right )\, 
\exp \left \{ -{1\over 2} \left [(\alpha^2 + \beta^2)\ln \epsilon_x 
+ (\alpha^2 - \beta^2)\ln 2x \right ]\right \} \, \, , &$\eta = 0$ ; \cr 
\sin \left ({\pi \over 2}\alpha \beta \right )\, 
\exp \left \{ -{1\over 2} \left [(\alpha^2 + \beta^2)\ln \epsilon_x 
- (\alpha^2 - \beta^2)\ln 2x \right ]\right \} \, \, , &$\eta = \infty $ . \cr } 
\neweq 
$$
The combinations (5..14,15) are the building blocks for 
the construction of both $J_\nu $ and $J_\nu^5$. Defining 
$$
Z_t (x,\epsilon_x ; \alpha ,\beta ) = Z_x^5 (x,\epsilon_x ; \alpha ,\beta ) = 
\left [\epsilon_x \beta K(x, \epsilon_x ;\alpha ,\beta ) \right ]^{-1} 
\quad , \neweq 
$$
and 
$$
Z_x (x,\epsilon_x ; \alpha ,\beta ) = Z_t^5 (x,\epsilon_x ; \alpha ,\beta ) = 
\left [\epsilon_x \alpha  K(x, \epsilon_x ;\alpha ,\beta ) \right ]^{-1} 
\quad , \neweq 
$$
we see that the limits in eqs.(5.8,9) give precisely (5.10). So, we are left 
with the equation of motion (5.11), which is satisfied provided that 
$$
\alpha - \beta = {g\over \pi } \quad . \neweq 
$$
Combining eq.(5.12) and eq.(5.19), we arrive at two families of solutions 
$$
\alpha_{1,2} = {g\over 2\pi } \pm \sqrt {{g^2\over 4\pi^2 } + (2k+1) } 
\quad , \qquad \beta_{1,2} = \alpha_{1,2} - {g\over \pi }  \quad .\neweq 
$$
parametrized by $k \in \Z$ with the constraint 
$$
2k+1 \geq -{g^2\over 4\pi^2 } \quad , \neweq 
$$
ensuring that $\alpha , \beta \in \R$. The freedom associated with 
$k$ is present also in the Thirring model on the plane. Lorentz invariance 
is however preserved there and if one requires the canonical value 
${1\over 2}$ for the Lorentz spin of $\Psi $ has, one finds $k=0$. 

The internal symmetries follow the pattern described in the previous 
section. It is convenient to normalize the vector and the chiral charges 
according to 
$$
\Q = - {1\over \alpha \sqrt \pi} Q \quad , \qquad 
\Q^5 = - {1\over \beta \sqrt \pi} \widetilde Q \quad .
\neweq 
$$
Under the vector boundary condition (5.4) 
(vector phase), $\Q$ generates 
the transformations (4.20) on $\Psi $ and $\overline \Psi$, 
whereas the action of $\Q^5$ is trivial. Accordingly, the physical 
representation in this case is reconstructed from the 
$\Q$-invariant correlation functions.  
If one imposes instead the axial boundary condition (5.5) 
(axial phase), $\Q^5$ transforms $\Psi_{{}_{L,R}}$ according to 
(4.23) and $\Q$ is trivial. Now the physical representation 
is recovered from the $\Q^5$-invariant correlators. 

Compared to the free Dirac equation on $\R \times \hl $, 
the impact of the duality transformation (2.57) on the above 
solutions of the Thirring model is slightly more involved. 
Taking into account that $\sc \leftrightarrow \dsc $ in (5.6,7) is 
equivalent to $\alpha \leftrightarrow \beta $ and 
in view of (5.19), one finds that (2.57) maps the vector phase 
with coupling $g$ in the axial phase with coupling $-g$ and vice versa. 

We turn now to the construction of 
a solution of the Thirring model on $\R \times \hl $, in which both 
vector and axial symmetry are nontrivially implemented. 
As in the free case, treated in the previous section, 
the idea is to adopt a superposition of two fields $\sc_1$ and 
$\sc_2$ obeying Neumann and 
Dirichlet boundary conditions respectively, in order to obtain 
$$
J_\nu (t,x) = - {1\over \sqrt {2\pi}}\left [\der_\nu \sc_1 (t,x) + 
\der_\nu \sc_2 (t,x) \right ] 
\quad , \neweq 
$$
$$
J_\nu^5 (t,x) = - {1\over \sqrt {2\pi}}\left [\der_\nu \dsc_1 (t,x) + 
\der_\nu \dsc_2 (t,x) \right ] 
\quad . \neweq 
$$
These currents satisfy the boundary condition 
$$
\lim_{x \downarrow 0}J_x (t,x) = 
U\left [\lim_{x \downarrow 0}J^5_x (t,x) \right ] U^{-1}  
\quad , \neweq 
$$ 
the operator $U$ being defined by eq.(4.33). Generalizing eqs.(4.29,30), 
we consider 
$$
\Psi_1 (t,x) = 
{\mu^{{1\over 4}(\alpha^2 + \beta^2)} \over \sqrt {2\pi}} 
:\exp \left \{i\sqrt {{\pi \over 2}} 
\left [ \alpha (\sc_1 + \sc_2 ) + \beta (\dsc_1 + \dsc_2 )\right ] \right \}: (t,x)   
\quad , \neweq 
$$
$$
\Psi_2 (t,x) = 
{\mu^{{1\over 4}(\alpha^2 + \beta^2)} \over \sqrt {2\pi}} 
:\exp \left \{i\sqrt {{\pi \over 2}} 
\left [ \alpha (\sc_1 + \sc_2 ) - \beta (\dsc_1 + \dsc_2 )\right ] \right \}: (t,x)   
\quad . \neweq 
$$
With this ansatz we have to solve the equation of motion 
$$
i(\gamma_t \der_t - \gamma_x \der_x)\Psi (t,x) = 
-{g\over \sqrt {2\pi }} : \left [\gamma_t \der_t (\sc_1 + \sc_2 ) - 
\gamma_x \der_x (\sc_1 + \sc_2 ) \right ] \Psi : (t,x)  \quad , \neweq 
$$
and to reproduce (5.23,24) by means of (5.8,9) with appropriate 
renormalization constants 
${\widetilde Z}_\nu (x,\epsilon_x ;\alpha ,\beta )$ 
and ${\widetilde Z}_\nu^5 (x,\epsilon_x ;\alpha ,\beta )$. 
A direct inspection shows that these requirements imply that 
$\alpha $ and $\beta $ are 
precisely given by eqs.(5.20,21). The renormalization constants 
turn out to be $x$-independent and read 
$$ 
{\widetilde Z}_t = {\widetilde Z}_x^5 = 
{\epsilon_x^{\, \, {1\over 2}(\alpha^2 + \beta^2) - 1}\over 
\beta \sin \left ({\pi \over 2}\alpha \beta \right )} 
\quad , \qquad  
{\widetilde Z}_x = {\widetilde Z}_t^5 = 
{\epsilon_x^{\, \, {1\over 2}(\alpha^2 + \beta^2) - 1}\over 
\alpha \sin \left ({\pi \over 2}\alpha \beta \right )} 
\quad . \neweq 
$$
The physical representation in this vector-axial 
symmetric phase is reconstructed from those correlation functions of 
the fields (5.26,27), which are both $\Q$ and $\Q^5$-invariant.  

Let us discuss finally the scaling 
properties of the obtained solutions, associated with different boundary 
conditions. These properties can be extracted from the behavior 
of the relative correlation functions in the physical representation. 
One finds that the dilatation 
group (2.46) is unitary implemented (3.22-24) with the following 
scale dimensions: 
$$
d_\Psi = \cases { {1\over 2}\alpha^2\, \, , & vector phase; \cr 
{1\over 2}\beta^2\, \, , & axial phase; \cr  
{1\over 4}(\alpha^2 + \beta^2 ) \, \, , & vector-axial phase. \cr } 
\neweq 
$$ 
At fixed coupling constant g, the dimension $d_\Psi $ takes 
different values in the above phases, except for the case 
$g^2 = (8k+4)\pi^2$ with $k\in \N$, in which $d_\Psi = k+1/2$ in 
all phases. 

Summarizing, we derived in this section the operator solutions of 
the massless Thirring model on $\R \times \hl $, associated with the 
boundary conditions (5.4,5,25). The basic properties of the 
corresponding phases were also analyzed.

\newchapt {Conclusions}

In the present paper we developed a bosonization procedure on the half 
line. Our starting point was to prove that the massless scalar field on 
$\R \times \hl $ and its dual, possess some definite locality properties, 
in spite of the breakdown of translation and Lorentz invariance. 
These properties provide the basic 
ingredient for constructing fermion, and more generally anyon fields 
in terms of bosonic ones. We described various boundary 
conditions on the fermions, formulated in terms of the vector and 
axial currents and implemented by means of linear boundary conditions 
on the bosonic field. As it can be expected on general grounds, 
symmetries are deeply influenced by the boundary conditions, which give rise 
to phases with different symmetry content. We investigated 
this aspect in detail and established the fundamental role of duality in 
connecting these phases. 

The characteristic features of bosonization on the half line 
have been illustrated on the concrete 
example of the massless Thirring model. We constructed 
operator solutions corresponding to different boundary conditions. 
Structural issues like renormalization of the current operator 
and the symmetry properties of the different phases were also discussed. 

The above investigation is a step towards the understanding of the 
abelian bosonization on the half line. A natural and 
quite attractive problem for the future is to extend our results to 
the non-abelian case.

\vfill\eject

\centerline {\bbf Appendix A} 
\bigskip 
\medskip

We describe here the Fock representation $\F$ of the algebra $\A$, defined 
by the commutation relations (2.15,16). The delicate point is the 
distribution $\left (k_+^{-1} \right )_{\mu_{{}_0}}$ 
in the right hand side of (2.16). What we need there is a solution $\gamma (k)$ 
of the equation 
$$
k\, \gamma (k) = \theta (k) \quad . \eqno(A.1) 
$$
The naive one $\gamma (k) = \theta (k) k^{-1}$ has a 
non-integrable singularity in $k=0$ and therefore, is not 
a distribution over the Schwartz test function 
space $\S(\R)$. In spite of this fact, $\theta (k) k^{-1}$ gives rise 
to a continuous linear functional on the subspace\break  
$\S_0(\R) = \{f\in \S(\R)\, :\, f(0) = 0 \}$ which, according to the Hahn-Banach 
theorem, can be extended to the whole $\S(\R)$. 
The result of this operation is a one-parameter family of extensions, 
which can be represented by [26] 
$$
\left (k_+^{-1} \right )_{\mu_{{}_0}}\equiv  
{d \over dk} \left [ \theta (k) \ln {k\over \mu_{{}_0} } \right ] 
\quad , \eqno(A.2) 
$$
where $\mu_{{}_0} > 0$ is a free scale parameter 
and the derivative is in the sense of distributions. 
By construction $\left (k_+^{-1} \right )_{\mu_{{}_0}}$ satisfies (A.1) 
and has support in $\overline \R_+ $. For 
$\varrho >0$ one finds the following 
quasi-homogeneous behavior 
$$
\left (\left (\varrho k\right )_+^{-1} \right )_{\mu_{{}_0}} = 
\varrho^{-1}\left (k_+^{-1} \right )_{\mu_{{}_0}} + 
\varrho^{-1} \delta (k) \ln \varrho 
\quad . \eqno(A.3) 
$$
In the terminology of [26] $\left (k_+^{-1} \right )_{\mu_{{}_0}} $ is 
therefore an associated (generalized) function of first order 
and degree $-1$. 

We turn now to the algebra $\A$. For the smeared fields 
$$
a(f) = \int_{-\infty }^\infty \mis \overline f(k)a(k) \quad , \qquad 
a^*(f) = \int_{-\infty }^\infty \mis f(k)a^*(k) \quad , 
\qquad f,g\in \S(\R) \quad , \eqno(A.8) 
$$
the commutator (2.16) takes the form 
$$
[a(f)\, ,\, a^*(g)] = \langle f,g\rangle_{{}_\S} 
\quad , \eqno(A.4) 
$$
with 
$$
\langle f,g\rangle_{{}_\S} \equiv \int_{-\infty }^\infty \mis 
\left (k_+^{-1} \right )_{\mu_{{}_0}} \overline f(k)g(k) 
\quad . \eqno(A.5) 
$$
Since the distribution $\left (k_+^{-1} \right )_{\mu_{{}_0}} $ 
is not positive definite, from (A.4,5) we deduce that the 
quantization requires indefinite metric. 
In order to enter the standard framework 
of indefinite metric quantization (see e.g.[27]), 
we have to construct a one-particle Hilbert space 
$\{\H^{(1)},\, \sform^{(1)}\}$, equipped with 
a nondegenerate jointly continuous sesquilinear form 
$\form^{(1)}$, which is related to $\form_{{}_\S}$ as follows: 
there exists an operator $I$ mapping $\S(\R)$ on a dense subspace 
of $\H^{(1)}$, such that 
$$
\langle If,Ig\rangle^{(1)} =  \langle f,g\rangle_{{}_\S} 
\quad \qquad \forall f,g\in \S(\R) \quad . \eqno(A.6) 
$$ 
Let us summarize briefly the construction. First we 
observe that the restriction $\form_{_{\S_0}}$ of $\form_{_{\S}}$ 
on $\S_0(\R)$ is positive definite. Therefore 
$\{\S_0(\R)\, \, \form_{{}_{\S_0}}\}$ is a pre-Hilbert space. 
The corresponding Hilbert space will be denoted by 
$\{\H_0\, \, \sform_{{}_0}\}$. Now, as a one-particle space we take 
$\H^{(1)}\equiv \H_0\oplus \C^2$ with the standard scalar 
product 
$$
(\chi_1,\chi_2)^{(1)} = (f_1,f_2)_{{}_0} + \overline a_1a_2 + \overline b_1b_2 
\quad , \eqno(A.7) 
$$
and the sesquilinear form 
$$
\langle \chi_1,\chi_2\rangle^{(1)} = (f_1,f_2)_{{}_0} + \overline a_1b_2 + 
\overline b_1a_2 
\quad , \eqno(A.8) 
$$
$\chi_i = (f_i,a_i,b_i)$ being arbitrary vectors in $\H^{(1)}$. 
Obviously 
$$
| \langle \chi_1,\chi_2\rangle^{(1)}|^2 \leq 
(\chi_1,\chi_1)^{(1)} (\chi_2,\chi_2)^{(1)} \quad , 
$$ 
showing the joint continuity of the form. 

{}For constructing $I$, we fix $h\in \S(\R)$ satisfying 
$$
h(0) = 1 \quad , \qquad \langle h,h\rangle_{{}_\S} = 0 \quad . \eqno(A.9) 
$$
One can adopt for instance 
$$
h(k) = \exp \left (-{k^2\over 2\mu^2_{{}_0}\e^{\gamma_{{}_E}}}\right ) \quad . 
\eqno(A.10) 
$$
Then, setting 
$$
 f_h(k) = f(k) - f(0)h(k) \quad , \eqno(A.11) 
$$
one can directly verify that the mapping 
$$
If = (f_h, \langle h,f\rangle_{{}_\S}, f(0)) \quad , \eqno(A.12) 
$$
satisfies all requirements. 
So, we are in position to apply the results of [27], which generalize
the standard free field quantization procedure to the case of indefinite metric. 
Skipping the details, the basic steps are the following. 
We set 
$\H^{(0)} \equiv \C$ and\break $\H^{(n)} \equiv \otimes_{k=1}^n \H^{(1)}$. 
Let 
$$
\V^{(n)} = \{ \chi_1\otimes \cdots \otimes \chi_n \, :\, \chi_i \in \H^{(1)} \} 
\eqno(A.13) 
$$
be the total set of decomposable vectors in $\H^{(n)}$. We define on 
$\V^{(n)}$ the form 
$$
\langle \chi \, ,\, \chi^\prime \rangle =
\langle \chi_1\, ,\, \chi_1^\prime \rangle^{(1)}  \cdots 
\langle \chi_n\, ,\, \chi_n^\prime \rangle^{(1)} 
\eqno(A.14) 
$$
and extend it by linearity and continuity to $\oplus_{n=0}^\infty \H^{(n)}$. 
Moreover, for any $f\in \S(\R)$ we introduce 
the operators $b^*(f)$ and $b(f)$, which act on $\V^{(n)}$ as follows: 
$$
b^*(f) \chi_1\otimes \cdots \otimes \chi_n = If \otimes 
\chi_1\otimes \cdots \otimes \chi_n \quad , \eqno(A.15) 
$$
$$
b(f) \chi_1\otimes \cdots \otimes \chi_n = \langle If,\chi_1\rangle^{(1)} 
\chi_2\otimes \cdots \otimes \chi_n \quad , \qquad n\geq 1 \quad , \eqno(A.16) 
$$
and $b(f)\H^{(0)} = 0$. The operators (A.15,16) admit unique extensions by 
linearity and continuity to the whole $\H^{(n)}$. 
Now, we are ready to define the Fock space $\{\F,\, \form \}$, 
the vacuum state $\Omega $ and the operators $a(f)$ and $a^*(f)$, 
we are looking for. We set 
$$
\F = \oplus_{n = 0}^\infty S_n \H^{(n)} \quad , \eqno(A.17) 
$$
$S_n$ being the symmetrization operator. The inner product, 
used in the body of the paper for the evaluation of expectation values, is 
precisely the restriction of $\form $ on $\F$. The finite particle 
subspace $\F_0 \subset \F$ consists of sequences of the type 
$\chi = \left (\chi^{(0)}, \chi^{(1)},...,\chi^{(n)},...\right )$ 
with $\chi^{(n)}\in S_n\H^{(n)}$ and $\chi^{(n)}=0$ for $n$ large enough. 
The vector representing the vacuum state reads 
$\Omega = (1,0,...,0,...)\in \F_0$. Finally, we define 
the creation and annihilation operators 
$$
a^*(f) = \sqrt {n+1}\, S_{n+1}\, b^*(f) \quad , \eqno(A.16) 
$$
$$
a(f) = \sqrt n \, b(f) \quad , \eqno(A.17) 
$$
restricted to each $S_n \H^{(n)}$ and then 
extended by linearity to $\F_0$. 
One may check easily that both $a(f)$ and  $a^*(f)$ vanish 
whenever the support of $f$ lies out of $\overline \R_+ $. 

This concludes the construction of the Fock representation of the 
commutation relations (2.15,16), used throughout the paper.

\vfill\eject

\centerline {\bbf Appendix B}

\bigskip 
\medskip

According to eq.(3.8), a generic correlation function of the 
anyon field (3.1) is expressed in terms of the functions 
$W_{ij}(t,x,y;\eta )$. The explicit form of $W_{ij}$ is as follows: 
$$
W_{ij}(t,x,y; \eta = 0) = 
$$
$$
{1\over 4}\biggl \{ 
(\alpha_i \alpha_j + \beta_i \beta_j ) \ln [-t^2 +(x-y)^2 + i\epsilon t] + 
(\alpha_i \alpha_j - \beta_i \beta_j ) \ln [-t^2 +(x+y)^2 + i\epsilon t] + 
$$
$$
(\alpha_i \beta_j + \alpha_j \beta_i ) 
\ln {t-(x-y) - i\epsilon \over t+(x-y) - i\epsilon } -  
(\alpha_i \beta_j - \alpha_j \beta_i ) 
\ln {t-(x+y) - i\epsilon \over t+(x+y) - i\epsilon } \biggr \}  \quad , \eqno(B.1) 
$$
\medskip 
$$
W_{ij}(t,x,y; \eta = \infty) = 
$$
$$
{1\over 4}\biggl \{ 
(\alpha_i \alpha_j + \beta_i \beta_j ) \ln [-t^2 +(x-y)^2 + i\epsilon t] -
(\alpha_i \alpha_j - \beta_i \beta_j ) \ln [-t^2 +(x+y)^2 + i\epsilon t] + 
$$
$$
(\alpha_i \beta_j + \alpha_j \beta_i ) 
\ln {t-(x-y) - i\epsilon \over t+(x-y) - i\epsilon } +  
(\alpha_i \beta_j - \alpha_j \beta_i ) 
\ln {t-(x+y) - i\epsilon \over t+(x+y) - i\epsilon } \biggr \}  \quad , \eqno(B.1) 
$$
\medskip
$$
W_{ij}(t,x,y; 0< \eta < \infty ) = W_{ij}(t,x,y; \eta = \infty) + 
$$
$$
{1\over 2}
\biggl \{ (\alpha_i \alpha_j - \beta_i \beta_j - \alpha_i \beta_j + \alpha_j \beta_i ) 
\e^{-\eta [t-(x+y)]}\, {\rm Ei} [\eta t - \eta (x+y) - i\epsilon ] + 
$$
$$
(\alpha_i \alpha_j - \beta_i \beta_j + \alpha_i \beta_j - \alpha_j \beta_i ) 
\e^{\eta [t+(x+y)]}\, {\rm Ei} [-\eta t - \eta (x+y) + i\epsilon ] \biggr \}
\quad . \eqno(B.3) 
$$

\vfill\eject 

\centerline {\bbf References} 
\bigskip 
\medskip 

\item {[1]} I. Affleck and A. W. W. Ludwig, Nucl. Phys. {\bf B352} (1991) 841. 

\item {[2]} C. L. Kane, M. P. A. Fisher, Phys. Rev. Lett {\bf 68} (1992) 1220. 

\item {[3]} P. Fendley, A. W. W. Ludwig and H. Saleur, Phys. Rev. Lett {\bf 75} 
(1995) 2196. 

\item {[4]} F. Lesage, H. Saleur and S. Skorik, Nucl. Phys. {\bf B474} (1996) 602.  

\item {[5]} A. J. Legget, S. Chakravarty, A. T. Dorsey, M. P. A. Fisher, 
A. Garg and W. Zwerger, Rev. Mod. Phys. {\bf 59} (1987) 1. 

\item {[6]} C. G. Callan and L. Thorlacius, Nucl. Phys. {\bf B329} (1990) 117. 

\item {[7]} C. G. Callan, I. R. Klebanov, A. W. W. Ludwig and 
J. M. Maldacena, Nucl. Phys. {\bf B422} (1994) 417. 

\item {[8]} D. E. Freed, Exact solutions for correlation functions in some 
(1+1)-D field theories with boundary, hep-th/9503065.  

\item {[9]} E. K. Sklyanin, J. Phys. {\bf A21} (1988) 2375. 

\item {[10]} S. Ghoshal and A. Zamolodchikov, Int. J. Mod. Phys. 
{\bf A9} (1994) 3841. 

\item {[11]} P. Bowcock, E. Corrigan, P. E. Dorey and 
R. H. Rietdijk, Nucl. Phys. {\bf B445} (1995) 469. 

\item {[12]} A. Liguori, M. Mintchev and L. Zhao, Boundary exchange algebras 
and scattering on the half line, hep-th/9607085. 

\item {[13]} J. A. Swieca, Fortschritte der Physik {\bf 25} (1977) 303. 

\item {[14]} S. Coleman, Phys. Rev. {\bf D11} (1975) 2088. 

\item {[15]} S. Mandelstam, Phys. Rev. {\bf D11} (1975) 3026. 

\item {[16]} E. Witten, Commun. Math. Phys. {\bf 92} (1984) 455. 

\item {[17]} P. Jordan and E. P. Wigner, Zeitschr. f\"ur Phys. 
{\bf 47} (1928) 631. 

\item {[18]} A. S. Wightman, in {\it Carg\`ese Lectures in Theoretical 
Physics}, (Gordon and Breach, New York, 1964). 

\item {[19]} E. Abdalla, M. C. B. Abdalla and K. D. Rothe, 
{\it Non-perturbative methods in two-dimensional quantum field theory}, 
(World Scientific, Singapore, 1991). 

\item {[20]} R. F. Streater and A. S. Wightman, {\it PCT, Spin and 
Statistics, and All That}, (Addison-Wesley, Redwood City, 1989).  

\item {[21]} A. Liguori and M. Mintchev, in preparation. 

\item {[22]} J. L. Cardy, Nucl. Phys. {\bf B240} (1984) 514. 

\item {[23]} A. L. Carey, S. N. M. Ruijsennars and J. D. Wright, 
Commun. Math. Phys. {\bf 99} (1985) 347. 

\item {[24]} K. Johnson, Nuovo Cim. {\bf 20} (1961) 773. 

\item {[25]} W. Thirring, Ann. of Phys. {\bf 3} (1958) 91. 

\item {[26]} I. M. Gel'fand and G. E. Shilov, 
{\it Generalized Functions}, v.1, (Academic Press, New York, 1964). 

\item {[27]} M. Mintchev, J. Phys. {\bf A13} (1980) 1841.

\vfill\eject 
\bye